\documentclass[aps,prl,twocolumn,amsmath,amssymb,superscriptaddress,notitlepage]{revtex4-2}
\usepackage{graphicx,amssymb,amsmath,amsfonts,empheq,braket}
\usepackage[dvipsnames]{xcolor}
\usepackage{dsfont}
\usepackage{dcolumn}
\usepackage{tikz}
\usepackage{textcomp}
\usepackage{float}
\usepackage{mwe}
\usepackage{ulem}
\usepackage{hyperref}
\hypersetup{
colorlinks = true,
linkcolor = [rgb]{0,0,0}, 
citecolor = [rgb]{0.13,0.55,0.13},
urlcolor = [rgb]{0.25, 0.41, 0.88}}

\usepackage{siunitx}
\usepackage{bm} % bold-italic

\renewcommand{\Re}{\operatorname{Re}}

\newcommand{\Tr}{\operatorname{Tr}}

\renewcommand{\r}{\boldsymbol{r}}
\newcommand{\q}{\boldsymbol{q}}
\newcommand{\m}{\boldsymbol{m}}
\newcommand{\lb}{\ell_B}

\newcommand{\eqnref}[1]{Eq.\,\eqref{#1}}
\newcommand{\figref}[1]{Fig.\,\ref{#1}}

\begin{document}
\title{Designing exciton-condensate Josephson junction in quantum Hall  heterostructures}
\author{Tianle Wang}
\affiliation{Department of Physics, University of California, Berkeley, CA 94720}
\affiliation{Materials Science Division, Lawrence Berkeley National Laboratory, Berkeley, California 94720, USA}
\author{Ruihua Fan}
\affiliation{Department of Physics, University of California, Berkeley, CA 94720}
\author{Zhehao Dai}
\affiliation{Department of Physics, University of California, Berkeley, CA 94720}
\affiliation{Department of Physics and Astronomy, University of Pittsburgh, Pittsburgh, PA 15260}
\author{Michael P. Zaletel}
\affiliation{Department of Physics, University of California, Berkeley, CA 94720}
\affiliation{Materials Science Division, Lawrence Berkeley National Laboratory, Berkeley, California 94720, USA}

\begin{abstract}
The exciton condensate (EC), a coherent state of electron-hole pairs, has been robustly realized in two-dimensional quantum Hall bilayer systems at integer fillings. However, direct experimental evidence for many of the remarkable signatures of phase coherence, such as an in-plane Josephson effect, has been lacking.  
In this work, we propose a gate-defined exciton-condensate Josephson junction suitable for demonstrating the Josephson effect in vdW heterostructures. The design is similar to the S-I-S superconducting Josephson junction but functions with a completely different microscopic mechanism: two exciton condensates are spatially separated by a gated region that is nearly layer-polarized, and the variation of layer pseudospin mediates a Josephson coupling sufficiently strong to have an observable effect. The Josephson coupling can be controlled by both the gate voltage and the magnetic field, and we show our design's high range of tunability and experimental feasibility with realistic parameters in vdW heterostructures.
\end{abstract}
\maketitle

The exciton condensate (EC) is a coherent state of electron-hole pairs predicted to exhibit many of the same phenomena as superconductivity, and its experimental demonstration remains a central pursuit in condensed matter physics~\cite{blatt1962BoseEinsteinCondensationExcitons,schafroth1954SuperconductivityChargedBoson,schafroth1954TheorySuperconductivity}. 
This phase was first realized in semiconductor quantum Hall bilayer systems at unit filling~\cite{eisenstein1990IndependentlyContactedTwodimensional,MacDonald2004nature,eisenstein2014ExcitonCondensationBilayer}, where excitons are formed by binding electrons in one layer with holes in the other.
Clear evidence of exciton condensation has been observed in terms of zero-resistance counterflow and perfect Coulomb drag~\cite{kellogg2004VanishingHallResistance,nandi201242ExcitonCondensation}, in analogy to the superflow in superconductivity. More recently, the EC has been realized in quantum-Hall van der Waals (vdW) heterostructures such as the graphene bilayers~\cite{liu2017Exp1QuantumHall, li2017Exp2ExcitonicSuperfluid, liu2022Exp3CrossoverStrongly} and transition metal dichalcogenides (TMD) bilayers~\cite{shi2022BilayerWSe2Natural}. Moreover, experiments in TMD bilayers have also identified ``exciton insulators'' at zero magnetic fields ~\cite{nguyen2023PerfectCoulombDrag, qi2023PerfectCoulombDrag}, where all electrons and holes strongly bind into excitons that possibly condense at low temperature, opening up another route to EC beyond quantum-Hall systems.

Despite this progress, EC systems have yet to demonstrate the diverse current-phase phenomena in superconductors, such as the Josephson effect, Shapiro steps, and Fraunhofer patterns. Although beautiful early studies of interlayer tunneling \cite{Spielman:2000zz,spielman200116ObservationLinearly} have identified a zero-bias conductance peak suggestive of the DC Josephson effect,  discrepancies with theoretical expectations, possibly due to disorders and resistive contacts, remains to be resolved \cite{eisenstein2014ExcitonCondensationBilayer}. Here, we propose the in-plane Josephson junction in exciton condensates~\cite{wen1996SidewaysTunnellingFractional,joglekar2005ExcitonicCondensateQuasiparticle,zhang2013ThermalTransportExcitoncondensate,zhang2021SustainingChargeneutralCharged} as an ideal platform to verify the excitonic Josephson effect.

\begin{figure}[t]
    \centering    
    \includegraphics[width=\linewidth]{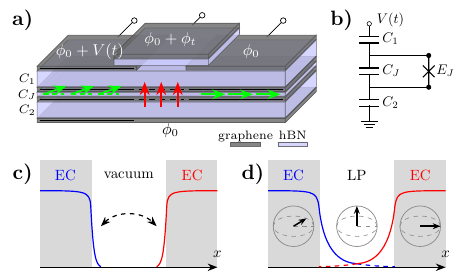}
    \caption{(a) Sketch of the proposed device for an exciton condensate Josephson junction. Dark and light slabs denote graphene (conductor) and hBN (insulator) respectively. The arrows denote the layer-pseudospin $\m$ orientation of the bilayer electron wavefunction at filling $\nu=1$ of the zeroth Landau level (see \eqnref{eq:EC wave function}): In-plane arrows (green) represent the exciton condensate (EC); vertical arrows (red) represent the layer-polarized (LP) integer quantum Hall state. (b) Equivalent lumped-element circuit of the device, with each element schematic drawn in (a): $C_J$ denotes the capacitance of electron bilayers, and $E_J$ denotes the EC Josephson energy from their coupling. $V(t)$ denotes the left-gate probe voltage capacitively coupled to the bilayer via $C_{1,2}$. (c-d) Comparison of two mechanisms of EC Josephson coupling: by tunneling excitons in vacuum junctions (c), and by in-plane canted pseudospin in layer-polarized junctions (d). The former becomes negligible at realistic junction spacings, while the latter extends much longer and is tunable by external fields.}
    \label{fig:Device}
\end{figure}

A major finding of our work is that the mechanism of excitonic Josephson coupling can differ significantly from that in superconductors, where it is mediated by electron tunneling across the junction. Due to the quenched kinetic energy of electrons in a strong magnetic field, such tunneling is greatly suppressed in quantum Hall systems --- in fact, in a clean system such tunneling is forbidden by the conservation of guiding center coordinate. Previous proposals for in-plane Josephson junctions thus relied on disorder \cite{wen1996SidewaysTunnellingFractional}, which is thought to be minimal in the gate-defined structures relevant to present vdW devices \cite{cohen2022UniversalChiralLuttinge}. 
Taking advantage of recent progress in fabricating gate-defined clean edges in vdW bilayers, we design a new type of exciton Josephson junction.
We show that the exciton Josephson coupling can be established from the canting of electron layer polarization in the quantum Hall ferromagnet \cite{moon199513SpontaneousInterlayer}. This mechanism does not rely on disorder and can be tuned via gate voltages, allowing the practical realization of a measurable EC Josephson effect in a clean quantum-Hall system.

The design of the junction is shown in \figref{fig:Device}(a):  two spatially separated ECs sandwich a junction region above which a split top-gate is used to apply a perpendicular displacement field, where the Josephson coupling is established and controlled. 
Much of this work is devoted to estimating the Josephson coupling in this setup.
We apply the Hartree-Fock approximation, or equivalently, a non-linear sigma model (NLsM) analysis to show that the Josephson coupling is highly tunable. 
Using this theory we propose a realistic design of an exciton-condensate Josephson junction embedded in an LC resonant circuit, where the Josephson effect can be measured directly through counterflow current or indirectly by measuring the AC impedance of the gates.
In particular, the oscillation frequency covers the microwave regime that is within the experimental reach.

\section{Device and measurement protocols}
\label{sec:device}

We consider a bilayer quantum Hall system with negligible interlayer hopping and gate-screened Coulomb interactions~\cite{WenZeeNeutralSuperfluid,Sondhi1993prb,moon199513SpontaneousInterlayer}. The layer degrees of freedom can be viewed as a pseudospin, and we use $\pm$ to label the two layers. There are two separately conserved charge $U(1)$ symmetries associated with the total electron number $N_{+} + N_{-}$ and the layer charge polarization $N_{+} - N_{-}$.
Throughout the bulk of our discussion, we focus on the unit total filling and assume that electrons stay within the zeroth Landau level (zLL).
Without interactions, there would be a huge pseudospin degeneracy at total filling one. It is well known that the exchange effect from Coulomb interaction completely polarizes the pseudospin~\cite{girvin199614MulticomponentQuantum}. Moreover, the finite layer separation favors the in-plane polarization, namely an equal-weight superposition of the two layers,
which spontaneously breaks the conservation of $N_{+} - N_{-}$. The system thus enters the EC phase.
Using $c_{+, k}$ ($c_{-, k}$) as the fermion operator for the $k$-th orbital of the zLL in the up (down) layer, the EC state can be well approximated by the following Slater determinant
\begin{equation}
\label{eq:EC wave function}
    \ket{\m(\theta,\varphi)}= \prod_k \left(\cos{\frac{\theta}{2}}\,c_{+, k}^\dag + e^{i\varphi}\sin{\frac{\theta}{2}}\,c_{-, k}^\dag\right) |0\rangle\,.
\end{equation}
Here $\theta \in [0,\pi]$ measures the layer polarization and can be adjusted by applying a perpendicular displacement field, while $\varphi$ characterizes the macroscopic EC phase which governs the Josephson effect.
We introduce a pseudospin variable $\bm{m} = (\sin\theta \cos\varphi, \sin\theta \sin\varphi, \cos\theta)$ to describe the charge polarization and interlayer phase coherence. The dynamics of the pseudospin is the only active degree of freedom within the variational manifold spanned by \eqnref{eq:EC wave function} and will play an important role in our later discussion.

We propose to realize an EC Josephson junction using the split top gate depicted in \figref{fig:Device}(a). Structures of this sort have been recently used, for example, to define tunable quantum point contacts \cite{ehrets2022Quantumpointcon,cohen2023NanoscaleElectrostaticControl} with a junction width of $50-$\SI{100}{nm}.
The bottom gate and the split top gates to the left and right are held at equal voltages, giving rise to layer-balanced ECs \cite{liu2017Exp1QuantumHall,li2017Exp2ExcitonicSuperfluid}; let this condition define the left / right top-gate voltage $\phi_0$.
A small additional voltage $\phi_t$ is then applied to the middle top gate, increasing both the chemical potential and the displacement field along a strip in the 2DEG below. Because of the large integer quantum Hall gap ($\Delta \sim \SI{10}{\meV}$)
at filling $\nu_T = 1$ \cite{liu2022Exp3CrossoverStrongly}, there is a range of $\phi_t$ where $\nu_T = 1$ remains unchanged, while the displacement field creates a layer-polarized strip between the two ECs which will define the ``tunneling barrier'' of our Josephson junction. We will discuss the relevant range of $\phi_t$  quantitatively in later sections.

The microscopic mechanism of Josephson coupling in our design differs from that of superconducting ones.
In a superconductor, the Josephson coupling originates from the single-electron hopping across the junction, a mechanism inapplicable to the quantum Hall EC in the absence of disorder.
This can be seen from zLL wave functions in the Landau gauge $\psi(x,y) \propto \exp{\big(ik_y y -(x - k_y \ell_B^2)^2/{2\ell_B^2}\big)}$, where we choose $y$ to run parallel to the junction.
Orbitals centered around different positions in the $x$ direction have different momentum $k_y$ in the $y$ direction. Thus, direct electron hopping across the junction violates momentum conservation and is forbidden in clean samples.
An alternative mechanism is electron-hole pair hopping across the junction, which conserves momentum and can be generated by the Coulomb interaction.
When the junction region has zero filling, this is the only source of the Josephson coupling (\figref{fig:Device}(c)). 
The amplitude of this process is proportional to the overlap between Landau orbitals from two sides, which, unfortunately, decays as a Gaussian in the width of the junction $L_J$, i.e. $E_J \sim e^{-L_J^2/2\lb^2}$.
In realistic experimental setups the junction width is several 10s of $\SI{}{nm}$, i.e. at least a few magnetic lengths, and we expect negligible Josephson coupling through this effect. 

The origin of the Josephson coupling proposed here instead originates when the junction is kept at the same integer filling as the EC, through the dynamics of the pseudospin.
Specifically, we choose appropriate gate voltages such that the pseudospin stays within the $xy$-plane in the two EC regions and is \emph{almost} polarized along the $z$-axis inside the junction.
We know from the theory of quantum-Hall ferromagnetism that the pseudospin must vary smoothly in space in order to minimize the Coulomb energy~\cite{girvin199614MulticomponentQuantum}.
Therefore, the pseudospin inside the junction will always have a small in-plane component coupling the exciton phase to the left / right of the junction, mediating a Josephson coupling (Fig.~\ref{fig:Device}(d))
\begin{equation}
    \Delta E = -E_J \cos (\varphi_L- \varphi_R)\,.
\end{equation}
We will show in the next section $E_J \sim e^{-L_J / \xi}$,
where $\xi$ is tunable by the displacement field, and can yield an experimentally feasible Josephson effect.

Before calculating $E_J$, we first review the lumped-element description of the device and its response. Similar to the Cooper pair box, we can interpret our device as an LC circuit where the two ECs act as a single capacitor and the junction as an inductor (\figref{fig:Device}(b))~\footnote{We note that a single capacitor $C_J$ can describe the layer polarization energy in both EC islands: for a fixed charge number in each layer, the layer polarization of two ECs will be equal and opposite, and $C_J$ will be equal to the series-connected sum of their capacitances.}. We can translate between the superconductor and the EC cases by mapping charge to layer polarization, voltage to layer voltage differences, and current to counter-flow current. This lumped-element description is applicable at low probe frequency $\nu \ll v/L_x$, with $v$ the velocity of EC phase variation and $L_x$ the size of EC islands; in this limit the EC phase gradient within the bulk is negligible (see App.\,\ref{app:lumped_element_validity} for an analysis without this assumption).

The total energy of the lumped-element device depends on EC layer polarization $N_\mathrm{x} = (N_{+} - N_{-})/2$ and EC phase difference $\varphi$ across the junction:
\begin{equation}
   E[N_\mathrm{x},\varphi] = \frac{d_\ell}{2d_g} N_\mathrm{x}eV(t) + \frac{e^2}{2C_J} N_\mathrm{x}^2 - E_J \cos\varphi.
   \label{eq:circuit}
\end{equation}
Here $d_\ell$ is the distance of bilayer separation, $d_g\gg d_\ell$ is the distance between the bilayers and the top/bottom gates, $V(t)$ is the extra voltage applied on the left top gate, and $C_J$ is the effective bilayer capacitance.  Curiously, $C_J$ is larger than its geometric capacitance due to the exchange interaction in EC states, proportional to $\ell_B d_\ell^{-2}$ instead of $d_\ell^{-1}\,$\cite{moon199513SpontaneousInterlayer}.
The device has a resonant frequency in response to $V(t)$
\begin{equation}
   \nu_J = e\sqrt{E_J/C_J}/h\,,
   \label{eq:Josephson_frequency}
\end{equation}
which can be probed by both conductance and capacitance measurements.
 
If layer-resolved contacts with the device can be established, we can connect the two layers of the right island and keep the left one floating (here $C_J$ only includes the left EC island). We can then read out the current oscillation $I_J(t)$ across the contacts driven by an AC displacement potential $V(t)=V_1\cos(2\pi\nu t)$:
\begin{equation}
\begin{aligned}
    I_J(t) &= \frac{dQ_J(t)}{dt} = \frac{e}{\hbar} E_J \sin\varphi_J(t)\\
    \frac{d\varphi_J(t)}{dt} &= \frac{e}{\hbar}\left(\frac{d_\ell}{2d_g}V(t) -  \frac{Q_J(t)}{C_J}\right)
\end{aligned} \label{eq:lumped_element_EOM}
\end{equation}
Here $Q_J=eN_\mathrm{x}$ and $\varphi_J=\varphi$ compared to \eqnref{eq:circuit}. For $I_J \ll e E_J/\hbar$ we may take $\sin\varphi_J\approx \varphi_J$ and derive the frequency dependence of its magnitude
\begin{equation}
    I_J(\nu)=\frac{2\pi\nu C_J}{1-(\nu/\nu_J)^2}\frac{d_\ell}{2d_g}V_1,
\end{equation}
which diverges as $\nu$ approaches $\nu_J$. The result is valid for $\frac{d_\ell}{d_g} eV_1\ll |\frac{1-(\nu/\nu_J)^2}{\nu/\nu_J}|h\nu_J$.

If good contacts are unavailable, such as in TMD devices, one can probe the AC-gate capacitance $\tilde{C}_g(\nu)=Q_g(\nu)/V(\nu)$ of the left gates sandwiching the EC island. 
Compared to the geometric capacitance $C_g\sim (2d_g)^{-1}$, we will observe a frequency dependent response singular at $\nu=\nu_J$:
\begin{equation}
\begin{aligned}
    \tilde{C}_g(\nu) = C_g\left(1+\frac{d_\ell^2}{4d_g^2}\frac{C_J/C_g}{1-(\nu/\nu_J)^2}\right)\\
    \sim C_g\left(1+\frac{\lb}{d_g}\frac{1}{1-(\nu/\nu_J)^2}\right)
\end{aligned}
\end{equation}
Remarkably, the Hartree-Fock result $C_J\sim \ell_B d_\ell^{-2}$ implies that the correction to the geometric capacitance $C_g$ is of order $\mathcal{O}(\ell_B/d_g)$, much larger than the naive expectation $\mathcal{O}(d_\ell/d_g)$.
At zero frequency, the effect is as if the condensate brings the top and bottom gates closer by a distance $\mathcal{O}(\ell_B)$.
Such a response can be measured, for example, by reflectometry measurements of the top gate~\cite{Vigneau:2022cud}.

%\section{Non-linear sigma model of excitonic Josephson junction}
\section{Pseudospin-mediated Josephson coupling}
\label{sec:NLSM}

To develop our intuition, we estimate the Josephson coupling $E_J$ within a simplified phenomenological model. This model can be derived by applying the Hartree-Fock approximation of the realistic system, including the details of the split-gate electrostatics, which is the focus of the next section (see App.\,\ref{app:Hartree-Fock} for details). In the case of unit filling and slow variations, the Hartree-Fock description is equivalent to a non-linear sigma model (NLsM) of the layer pseudospin $\bm{m}$, which is the only remaining degree of freedom in question.
Using $L_y$ as the length along the $y$ direction and $L_J$ as the junction width, and assuming translation invariance along the $y$ direction and slow spatial variation of $\bm{m}$, the minimal action can be written as~\cite{Sondhi1993prb,moon199513SpontaneousInterlayer}
\begin{equation}
\begin{aligned}
    \frac{E[\bm{m}]}{L_y} = \int \mathrm{d}x & \ \frac{\rho_{xy}}{2}\left[(\partial_x m_x)^2+(\partial_x m_y)^2\right]  \\
    & \quad + \frac{\rho_z}{2}(\partial_x m_z)^2 + \frac{\beta}{2} m_z^2-\Delta(x) m_z.
\end{aligned}
\label{eq:LandauGinzburg}
\end{equation}
Physically, $\rho_{xy}$ and $\rho_z$ are the stiffness of in-plane and out-of-plane pseudospin fluctuation, $\beta$ is the capacitance energy per unit area, and  $\Delta(x)$ is a spatially dependent layer-polarizing potential. The first four terms are determined by Coulomb interaction, and the last term depends on the displacement field applied within the junction. 
For uniform $\bm{m}$ within the EC island, this NLsM model can reproduce the lumped-element description \eqnref{eq:circuit} with $C_J = \frac{e^2 A}{(4\pi\lb^2)^2\beta}$ (see App.~\ref{app:lumped_element_validity} for details).

\begin{figure}
\centering   \includegraphics[width=0.85\linewidth]{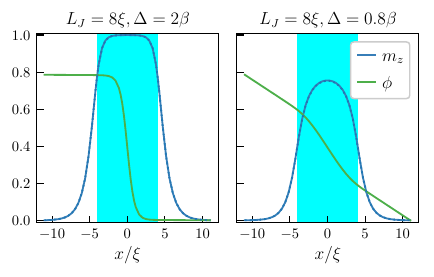}
    \caption{Comparison between weakly and strongly linked Josephson junction, with phase difference $(\varphi_L,\varphi_R)=(\pi/4,0)$ controlled by the layer-polarizing potential $\Delta(x)=0$ outside the junction and $\Delta(x)=\Delta$ within. In weakly linked junctions ($\Delta = 2 \beta$ is large), $\varphi(x)$ remains constant in each island and sharply changes within the junction; In strongly linked junctions (smaller $\Delta = 0.8 \beta$), $\varphi(x)$ varies smoothly from $\varphi_L$ to $\varphi_R$ across the junction.}
    \label{fig:JJ_comparison}
\end{figure}

We can determine the pseudospin configuration by minimizing \eqnref{eq:LandauGinzburg} with a given boundary condition.
Specifically, we fix the pseudospins at the two ends of the device to lie in the $xy$-plane with a relative angle $\varphi$, corresponding to two ECs with different phases.
For simplicity we choose $\Delta(x)$ to be a step function that takes a nonzero value $\Delta$ inside the junction and zero outside, and numerically solve for $\bm{m}$. 
As shown in \figref{fig:JJ_comparison}, depending on the ratio $\Delta / \beta$, two distinct behaviours can arise. 
For a weak polarizing potential $|\Delta| < \beta$, the junction itself is still an exciton condensate ($m_z < 1$), albeit with a smaller phase stiffness, across which the EC phase varies smoothly (\figref{fig:JJ_comparison}(b)).
Thus the Josephson coupling is too strong to identify a distinct junction from the EC bulks.
When the electric potential is sufficiently strong, $|\Delta|> \beta$, the pseudospin within the junction quickly polarizes to $m_z = 1$, and the EC phase exhibits a sharp jump at its center (\figref{fig:JJ_comparison}(a)). In this case the junction is more properly defined, and as we will show below, the coupling becomes exponentially small in the width of the junction. 

Leveraging the numerical results, we can give an analytical estimate of the Josephson energy $E_J$ as follows. 
When $\Delta > \beta$, we expect an almost polarized pseudospin $\bm{m}\simeq(0,0,1)$ deep inside the junction, and thus expand \eqnref{eq:LandauGinzburg} to the leading order of $m_x$ and $m_y$:
\begin{equation}
\begin{aligned}
    \frac{E[\bm{m}]}{L_y} \simeq \int dx\ \frac{\rho_{xy}}{2} & \big((\partial_x m_x)^2 + (\partial_x m_y)^2 \big) \\
    &+ \frac{\Delta - \beta}{2} (m_x^2+m_y^2)
\end{aligned}
    \label{eq:appquadE}
\end{equation}
where the $\rho_z$ term appears at the quartic order and is ignored.
The corresponding equations of motion are 
\begin{equation}
    \partial_x^2 m_\alpha = \frac{\Delta - \beta}{\rho_{xy}}m_\alpha\,,\quad \alpha = x,y\,.
    \label{eq:applinearEOM}
\end{equation}
Notably, the fluctuation of the pseudospin is governed by a newly emergent length scale
\begin{equation}
    \xi = \sqrt{\frac{\rho_{xy}}{\Delta - \beta}}\,.
\end{equation}
We need to specify the boundary conditions to fully determine the solutions.
Near the junction boundary, the approximation \eqref{eq:appquadE} is no longer accurate, but we still expect it to give a reasonable estimate of the Josephson coupling.
Accordingly, we choose the boundary condition $(m_x,m_y) = (\cos\varphi, \sin\varphi)$ at $x = -L_J/2$ and $(m_x,m_y) = (1, 0)$ at $x = L_J/2$.
In the long-junction limit $L_J\gg\xi$, we have
\begin{equation}
    \vec{m} = \vec{m}_L e^{-\frac{L_J/2 + x}{\xi}}
    + \vec{m}_R e^{-\frac{L_J/2 - x}{\xi}}
    \label{eq:applinearSol},
\end{equation}
where 
\begin{equation}
\begin{aligned}
    \vec{m}_L \simeq& (\cos\varphi - e^{-L_J/\xi}, \sin\varphi), \\
    \vec{m}_R \simeq& (1-e^{-L_J/\xi}\cos\varphi, -e^{-L_J/\xi}\sin\varphi)
\end{aligned}
\end{equation}
As a result, when the two ECs have a phase different $\varphi$ the energy of the junction is estimated as
\begin{equation}
    E \simeq -E_J \cos\varphi + \text{const.}
\end{equation}
where the Josephson energy reads
\begin{equation}
    E_J \simeq 2L_y\xi (\Delta-\beta) e^{-L_J/\xi}
    \label{eq:Josephson energy}
\end{equation}
Plugging the expressions of the parameters derived in App.~\ref{app:Hartree-Fock}, we have $E_J \sim \frac{e^2}{\epsilon \ell_B} \frac{d_\ell L_y}{\ell_B^2} e^{-L_J / \xi}$.
Numerical computation of the complete NLsM, including proper gate electrostatics, yields a Josephson coupling of the same order of magnitude as our analytic estimation.

The pseudospin-mediated Josephson coupling in \eqref{eq:Josephson energy} has two significant differences from that generated by the exciton hopping.
It decays single-exponentially with respect to $L_J$,  and can naturally be much larger than the case of a $\nu = 0$ junction, where the tunneling decays as a Gaussian in the clean case. 
More importantly, the correlation length $\xi$ is not solely tied to the magnetic length but can also be tuned continuously by the displacement field, which is the underlying reason for the high tunability of the proposed device.

\section{Optimal regime of Josephson effect}
\label{sec:tuning}

We now compute the Josephson energy $E_J$ for the graphene-hBN-graphene heterostructure as a concrete example, and estimate the optimal regime for observing the Josephson effect.
As shown in \figref{fig:Device}(a) there are three tuning parameters: the bottom gate voltage $\phi_0$, the middle top gate voltage $\phi_t$, and the magnetic field $B$.
The electrons in the zLL of the graphene heterostructure have eight flavors: two layers $\pm$, two valleys $K,K'$, and two spins $\uparrow/\downarrow$.
We assume an hBN sublattice splitting and Zeeman energy that can fully gap out the valley and spin sectors \cite{Goerbig:2010review}, thus we keep only one valley and one spin per layer in the low-energy manifold. The Hamiltonian includes gate-screened Coulomb interactions and electrostatic potentials from $\phi_0$ and $\phi_t$:
\begin{equation}
\begin{aligned}
    H = \sum_{a,b} \int d\bm{r}_1 d\bm{r}_2 & V_{ab}(\bm{r}_1 - \bm{r}_2) n_a(\bm{r}_1) n_b(\bm{r}_2) \\
    & + \sum_{a} \int d\bm{r} U_a(\bm{r}) n_a(\bm{r})
\end{aligned}
\end{equation}
where $n_a$ is the electron density in layer $a=\pm$. Both $V_{ab}(\r)$ and $U_a(\bm{r})$ are obtained by solving the Poisson equation within the split-gate device geometry (see App.\,\ref{app:NLsM_parameters}), assuming translational invariance along the junction.

For each set of parameters, we solve the Hamiltonian within the Hartree-Fock approximation (see App.\,\ref{app:Hartree-Fock}), which at unit filling (with properly chosen $\phi_0$) is approximately equivalent to solving the NLsM in \eqnref{eq:LandauGinzburg}.
To extract $E_J$, we work on a torus geometry and thread $\pm \varphi/2$ flux through the $x$-cycle of the top/bottom layer, effectively introducing a twist in the EC phase across the junction. Using HF to obtain the resulting energy $E[\varphi]$, for weak junctions the data is well fit to the form $E_J \cos\varphi$ from which we infer $E_J$. Meanwhile, we determine $C_J = \frac{e^2 A}{(4\pi\lb^2)^2\beta}$ by calculating $\beta$ analytically within Hartree-Fock, which to leading order is given by
\begin{equation}
    \beta\approx \frac{d_\ell^2}{8\sqrt{2\pi}}\lb^{-4}E_c,\ E_c=\frac{e^2}{4\pi\epsilon\lb}.
    \label{eq:beta_solution}
\end{equation}
We finally compute the resonant frequency $\nu_J = e \sqrt{E_J/C_J}/h$ featured in the proposed measurements.

\begin{figure}[tb]
    \centering
    \includegraphics[width=\linewidth]{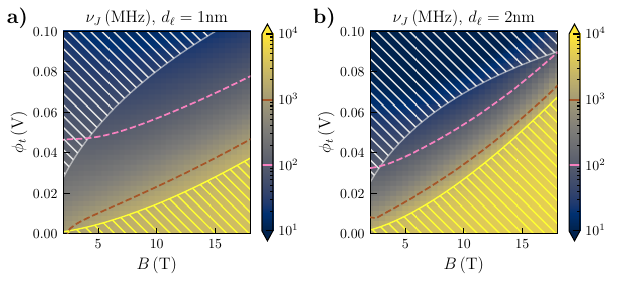}
    
    \caption{Resonant frequency $\nu_J(B,\phi_t)$ for $B\in(2,18)\,\si{T}$ and $\phi_t\in(0,0.1)\si{V}$. Device parameters: gate distance $d_g = \SI{25}{nm}$, junction width $L_J=\SI{250}{nm}$, island area size $L_x=L_y=\SI{5}{\micro\meter}$, dielectric constant $\epsilon = 4\epsilon_0$, with two different layer distances: (a) $d_\ell=\SI{1}{nm}$ and (b) $d_\ell=\SI{2}{nm}$. The white and yellow shaded regions correspond to the upper and lower bound in \eqnref{eq:phi_min_max}. The pink and red dashed lines are the contours of $\SI{100}{MHz}$ and $\SI{1000}{MHz}$.}
    \label{fig:vJ_phase_diagram}
\end{figure}

\figref{fig:vJ_phase_diagram} shows $\nu_J(B,\phi_t)$ as a function of $\phi_t$ and $B$ with a set of realistic device parameters. The desired junction is realized only within the unshaded region, which we will clarify shortly. Regardless, this region already realizes $\nu_J$ across three orders of magnitude in frequency,  and between the pink and red lines, which covers the microwave regime $\nu_J\in(100,1000)\,\si{MHz}$, is where the proposed effect is most accessible in the experiment.

The yellow/white-shaded regions in the above phase diagram denote the realistic constraints on our tuning parameters. For given $B$, $\phi_t$ is lower bounded by the condition $\Delta > \beta$ required to fully layer-polarize the junction, so that the weak link is sharply defined and the resulting $\nu_J$ is not too large. On the other hand, $\phi_t$ is upper bounded by our requirement of $\nu=1$ total filling: a large $\phi_t$ will raise the chemical potential of the junction beyond the charge gap of the EC, putting it above or below the unit filling such that it no longer serves a good barrier for excitons. Note that if split gates can be fabricated (and aligned) both above and below the junction, the middle gates can be asymmetrically biased to apply a layer-polarizing potential without a change in chemical potential, removing this upper-bound constraint and allowing further tunability.

As is detailed in App.~\ref{app:charge gap}, in the limit $d_\ell/d_g \ll 1$ we have the following range of $\phi_t$ for optimally chosen $\phi_0$
\begin{equation}
    8\pi\lb^2 d_g\beta/d_\ell < e\phi_t < 2V_{+-}^F\,,
    \label{eq:phi_min_max}
\end{equation}
where $\beta$ is given in \eqnref{eq:beta_solution}, and $V_{+-}^F\approx (\sqrt{\pi/2}-d_\ell)E_c$ is the exchange (Fock) Coulomb energy setting the EC charge gap. Here we ignore the hBN potential and Zeeman energy; including these terms only makes the conditions easier to satisfy. 
For small $d_\ell$, the lower bound is linear in $d_\ell$ and the upper bound is independent in $d_\ell$, giving a parametrically large range of $\phi_t$.
However, the Hartree-Fock approximation tends to overestimate the charge gap, which can lead to a smaller allowed range of $\phi_t$ than the simple estimate here.

\section{Discussion}

In our proposal, the Josephson coupling is mediated by the layer pseudospin fluctuation and is highly tunable by electrostatic gates. Our microscopic analysis focuses on the interlayer EC in quantum Hall graphene-hBN-graphene heterostructures at unit filling, but the proposal directly applies to other quantum-Hall bilayer van der Waals heterostructures \cite{shi2022BilayerWSe2Natural}.
Furthermore, the pseudospin-mediated coupling can be established between the junction barrier and EC islands at different electron fillings, e.g. between $\nu=1$ EC and $\nu=2$ junction, as long as the junction region is not a pseudospin singlet.
Exploring what happens when the middle region is at fractional fillings is theoretically and experimentally interesting.
Notably, bilayer EC can potentially be realized at zero magnetic fields \cite{nguyen2023PerfectCoulombDrag,qi2023PerfectCoulombDrag}. Despite the absence of Landau level and pseudospin description, we can consider the strength of Josephson coupling to be tunable by the exciton density inside the junction. We leave the detailed analysis of such devices to future work.

Another interesting tuning knob in vdW realizations is the twist angle between the electron bilayers. In vdW materials where the Landau levels emerge from the finite-momentum $K$ points (as in graphene), a relative twisting of the two layers can offset their electron momentum and prevent electrons from tunneling into the other layer. This can suppress the effect of interlayer tunneling, which has been shown to prevent the observation of the true Josephson effect for in-plane junctions ~\cite{Spielman:2000zz,stern200149TheoryInterlayer,joglekar200151ThereDc,rossi200563InterlayerTransport,park2006CoherentTunnelingExciton,hsu2015FractionalSolitonsExcitonic}. On the other hand, one can also enhance interlayer tunneling by careful alignment of electron bilayers, allowing invetigation of its effect on the Josephson junction.

Beyond demonstrating macroscopic phase coherence, the EC Josephson junction could offer utility as a circuit element within quantum-Hall setups, in which superconductivity cannot survive under large magnetic fields. For example, it might provide a high-Q resonator for dispersive readout of the charge state of an FQH state in a layer below. To this end, it would be useful for future theoretical work to estimate the factors limiting the quality of the junction, such as the dissipative transport of gapped merons parallel to the junction and the effect of thermal fluctuation on phase coherence.

\section{Acknowledgement}
We acknowledge Jiaqi Cai, Bertrand Halperin, Philip Kim, Ruishi Qi, Feng Wang, Taige Wang, Andrea Young,
and in particular Cory Dean and Allan Macdonald for enlightening discussions. 
We thank Zeyu Hao for providing useful feedback on the manuscript.
T.W. is supported by the U.S. Department
of Energy, Office of Science, Office of Basic Energy Sciences, Materials Sciences and Engineering Division under
Contract No. DE-AC02-05-CH11231 (Theory of Materials program KC2301). M.Z. and Z.D. are primarily supported by the U.S. Department of Energy, Office of Science, Basic Energy Sciences, under Early Career Award No. DE-SC0022716. R.F. is supported by the Gordon and Betty Moore Foundation (Grant GBMF8688).

\bibliography{bibliography}

\onecolumngrid
\newpage
\appendix
\setcounter{secnumdepth}{2}

\section{Derivation of non-linear sigma model with Hartree-Fock approximation}
\label{app:Hartree-Fock}

In this section, we derive the non-linear sigma model (NLsM) by using the Hartree-Fock approximation~\cite{moon199513SpontaneousInterlayer}. It also provides an estimate of the coefficients in the NLsM in terms of microscopic parameters, which is useful for calculating the Josephson coupling in realistic devices.
We begin with writing down the Hartree-Fock Hamiltonian in most general situations, multi-flavor electrons in the lowest Landau level (LLL).
Then, we derive the NLsM for the specific case, the bilayer quantum Hall problem with spinless fermion.

\subsection{General formulation of the Hartree-Fock Hamiltonian}
\label{sec:HF Hamiltonian}

Consider a two-dimensional electron gas with various flavors labeled by $a$, which can be layers, spins, etc.
We assume a generic two-body interaction $V_{ab}(\r)$ and an external potential $U_a(\r)$.
The Hamiltonian after the LLL projection is
\begin{equation}
    H = \frac{1}{2}\sum_{a,b} \int V_{ab} (\r_1 - \r_2) c_a^\dag(\r_1) c_b^\dag(\r_2) c_b(\r_2) c_a(\r_1) d\r_1 d\r_2 + \sum_{a} \int U_a(\r) c^\dag_{a}(\r) c_{a}(\r) d\r\,,
\end{equation}
where $c_a(\r)$ is the fermion annihilation operator projected onto the LLL.
We choose a periodic boundary condition along the $y$ direction and adopt the Landau gauge $A_x = 0$, $A_y = -Bx$, and have
\begin{equation}
    c_a(\r) = \frac{1}{\sqrt{L_y}} \sum_k \varphi_{k}(x) e^{iky} c_{k,a}\,,\quad \varphi_k(x) = \Big( \frac{1}{\pi \ell_B^2} \Big)^{1/4} e^{-\frac{(x - k \ell_B^2)^2}{2\ell_B^2}}
\end{equation}
where $k$ is the momentum in the $y$ direction and also controls the center of the wave function $\varphi_k(x)$ in the $x$ direction.
It is more convenient to work in the momentum space. Therefore, let us introduce $A = L_x L_y$ as the area of the 2D system and 
$
    V_{ab}(\r) = \frac{1}{A} \sum_{\q} \tilde{V}_{ab}(\q) e^{-i\q \cdot \r}
$
as the two-body interaction in the momentum space.
The interacting Hamiltonian can be then rewritten as
\begin{equation}
    H_{\text{int}} = \frac{1}{2A} \sum_{a,b} \sum_{\q, k, k'} \tilde{V}_{ab}(\q) F(\q) F(-\q) e^{iq_x(k' - k)\ell_B^2} c_{k-\frac{q_y}{2},a}^\dag c_{k'+\frac{q_y}{2},b}^\dag c_{k'-\frac{q_y}{2},b} c_{k+\frac{q_y}{2},a}
\end{equation}
where $F(\q) = e^{-|\q|^2\lb^2/4}$ is the LLL form factor.

In the Hartree-Fock approximation, we assume that the ground state and the low-energy excited states can be well approximated by Slater determinants. 
Within this variational manifold, the interacting Hamiltonian is decoupled into a Hartree and Fock term $H_{\text{int}} = H_{\text{Hartree}} + H_{\text{Fock}}$. 
The Hartree term comes from the direct interaction
\begin{equation}
\label{eq:H Hartree original}
    H_{\text{Hartree}} = \frac{1}{A} \sum_{a,b} \sum_{\q, k, k'} \tilde{V}_{ab}(\q) F(\q) F(-\q) e^{iq_x(k' - k)\ell_B^2} c_{k-\frac{q_y}{2},a}^\dag c_{k+\frac{q_y}{2},a} \braket{c_{k'+\frac{q_y}{2},b}^\dag c_{k'-\frac{q_y}{2},b}} 
\end{equation}
The Fock term comes from the exchange interaction
\begin{equation}    
\label{eq:H Fock original}
    H_{\text{Fock}} = -\frac{1}{A} \sum_{a,b} \sum_{\q, k, k'} \tilde{V}_{ab}(\q) F(\q) F(-\q) e^{iq_x(k' - k)\ell_B^2} c_{k-\frac{q_y}{2},a}^\dag c_{k'-\frac{q_y}{2},b} \braket{c_{k'+\frac{q_y}{2},b}^\dag c_{k+\frac{q_y}{2},a}} 
\end{equation}
where the minus sign comes from the fermi statistics.

It will help both the analytical and numerical manipulation if we can write the Hartree and Fock term in a similar form and treat them on an equal footing.
Let us introduce
\begin{equation}
\label{eq:nab}
    n_{ab}(\q)=\sum_k e^{-iq_x k\lb^2} c^\dag_{k-q_y/2,a}c_{k+q_y/2,b}\,,
\end{equation}
which characterizes the charge fluctuation in the momentum space. Note that so-defined $n_{ab}(\q)$ is different from the naive Fourier transformation of the projected density operator $c_a^\dag (\r) c_b(\r)$.
The Hartree term becomes
\begin{equation}
\label{eq:Hartree Hamiltonian}
    H_{\text{Hartree}} = \frac{1}{A} \sum_{a,b} \sum_{\q} \tilde{V}^{\text{Hartree}}_{ab}(\q) n_{aa}(\q) \braket{n_{bb}(-\q)} \,,\quad \tilde{V}^{\text{Hartree}}_{ab}(\q) = \tilde{V}_{ab}(\q) F(\q) F(-\q)\,.
\end{equation} 
For the Fock term, we first realize that the four fermion operators can be rewritten via $n_{ab}$ by manipulating the momenta as follows
\begin{equation*}
\begin{aligned}
    & \quad\  \sum_{k,k'} e^{iq_x(k' - k)\ell_B^2} c_{k-\frac{q_y}{2},a}^\dag c_{k'-\frac{q_y}{2},b} \braket{c_{k'+\frac{q_y}{2},b}^\dag c_{k+\frac{q_y}{2},a}} \\
    & = \sum_{k_1,p_y} e^{iq_x p_y \ell_B^2} c_{k_1-\frac{p_y}{2}-q_y,a}^\dag c_{k_1 + \frac{p_y}{2}-q_y,b} \langle c_{k_1+\frac{p_y}{2},b}^\dag c_{k_1-\frac{p_y}{2},a} \rangle\,,\quad \big(k = k_1 - \frac{p_y}{2} - \frac{q_y}{2}\,,\, k' = k_1 + \frac{p_y}{2} - \frac{q_y}{2} \big) \\
    & = \frac{1}{N_\phi} \sum_{k_1,k_2, \bm{p}} e^{iq_x p_y \ell_B^2 - i(k_1 - k_2) p_x \ell_B^2} c_{k_1-\frac{p_y}{2}-q_y,a}^\dag c_{k_1+\frac{p_y}{2}-q_y,b} \langle c_{k_2+\frac{p_y}{2},b}^\dag c_{k_2-\frac{p_y}{2},a}\rangle, \quad (N_\phi=A/2\pi \lb^2)\\
    & = \frac{1}{N_\phi} \sum_{\bm{p}} e^{i (q_x p_y - q_y p_x) \ell_B^2} n_{ab}(\bm{p}) \braket{n_{ba}(-\bm{p})}
\end{aligned}
\end{equation*}
After renaming the dummy variables we have
\begin{equation}
\label{eq:Fock Hamiltonian}
    H_{\text{Fock}} = -\frac{1}{A} \sum_{a,b} \sum_{\q} \tilde{V}^{\text{Fock}}_{ab}(\q) n_{ab}(\q) \braket{n_{ba}(-\q)}\,,\quad
    \tilde{V}^{\text{Fock}}_{ab}(\q) = \frac{1}{N_\phi} \sum_{\q'} \tilde{V}_{ab}(\q') F(\q') F(-\q') e^{i(q_x' q_y - q_y' q_x)\ell_B^2}
\end{equation}
This way the Hartree \eqref{eq:Hartree Hamiltonian} and Fock Hamiltonian \eqref{eq:Fock Hamiltonian} are in the same form and we can write the total Hartree-Fock Hamiltonian as 
\begin{equation}
    H_{\text{HF}} = \frac{1}{A} \sum_{a,b} \sum_{\bm{q}} \tilde{V}_{ab}^{\text{HF}}(\bm{q}) n_{ab}(\q) \,,\quad 
    \tilde{V}_{ab}^{\text{HF}}(\bm{q}) = \delta_{a,b} \sum_{c} \tilde{V}^{\text{Hartree}}_{ac}(\bm{q}) \braket{n_{cc}(-\bm{q})} - \tilde{V}_{ab}^{\text{Fock}}(\bm{q}) \braket{n_{ba}(-\bm{q})}
\end{equation}
with which one can have a more efficient numerical implementation than using their original form \eqref{eq:H Hartree original} and \eqref{eq:H Fock original}.

We assume that the ground state is translationally invariant along the $y$ direction but not necessarily along the $x$ direction. Then the fermion two-point function is
\begin{equation}
    \langle c^\dag_{k,b}c_{k',a}\rangle = \delta_{k,k'} P_{ab}(k)
\end{equation}
where $P(k)^2 = P(k)$ is a Hermitian spectral projector.
In the Landau gauge $P_{ab}(k)$ captures the charge fluctuation around $x = k \ell_B^2$. Accordingly, we can define the real-space interaction as
\begin{equation}
    V_{ab}^{X}(x) = \frac{1}{L_x}\sum_{q_x} \tilde{V}_{ab}^X(q_x) e^{i x q_x}\,,\quad X = \text{Hartree or Fock}
\end{equation}
and write the Hartree and Fock energy as (note the extra factor of $2$ comparing with the expression of the Hartree Fock Hamiltonians)
\begin{equation}
\label{eq:Hartree Fock energy real space space}
\begin{aligned}
    E_{\text{Hartree}}[P] &= \frac{1}{2 L_y}\sum_{a,b} \sum_{x,x'} V_{ab}^{\text{Hartree}}(x - x') P_{aa}(x) P_{bb}(x') \\
    E_{\text{Fock}}[P] &= \frac{1}{2 L_y}\sum_{a,b} \sum_{x,x'} V_{ab}^{\text{Fock}}(x-x') P_{ab}(x) P_{ba}(x')
\end{aligned}
\end{equation}
where we write $P_{ab}(x)$ in place of $P_{ab}(k = x \ell_B^{-2})$ for the clarity of the expresion.
Equivalently, we can define the charge fluctuation in the momentum space $\tilde{P}(q_x) = \sum_{k} e^{-i k q_x \ell_B^2} P (k)$ and have
\begin{equation}
\label{eq:Hartree Fock energy momentum space}
\begin{aligned}
    E_{\text{Hartree}}[\tilde{P}] &= \frac{1}{2A} \sum_{a,b} \sum_{q_x} \tilde{V}_{ab}^{\text{Hartree}}(q_x) \tilde{P}_{aa}(q_x) \tilde{P}_{bb}(-q_x) \\
    E_{\text{Fock}}[\tilde{P}] &= -\frac{1}{2A} \sum_{a,b} \sum_{q_x} \tilde{V}_{ab}^{\text{Fock}}(q_x) \tilde{P}_{ab}(q_x) \tilde{P}_{ba}(-q_x)
\end{aligned}
\end{equation}
When analyzing situations where the configuration is slowly varying along the $x$-direction, Eq.~\eqref{eq:Hartree Fock energy real space space} or \eqref{eq:Hartree Fock energy momentum space} will be a convenient starting point for approximations.

\subsection{Non-linear sigma model}

In this section, we derive the non-linear sigma model (NLsM) used in the main text. 
Our derivation shares the same spirit as the Sec. IV C in Ref.~\cite{moon199513SpontaneousInterlayer} and has the advantage of being easily generalizable.

Consider the bilayer quantum Hall problem where the fermions have two flavors $a = \pm$ corresponding to the two layers. 
One can consider a global $SU(2)$ transformation that rotates the layer-pseudospin of the electrons, which will become a symmetry of the Hamiltonian when the interaction is fully isotropic.
Here we only assume a symmetric interaction, i.e., $V_{++} = V_{--}$ and $V_{+-} = V_{-+}$ and organize the Hartree-Fock energy \eqref{eq:Hartree Fock energy momentum space} by the $SU(2)$ symmetry.
The Hartree energy can be decomposed into an $SU(2)$-symmetric part depending only on the total density and another $SU(2)$-symmetry-breaking part that comes from layer charge polarization and the anisotropy of the interaction
\begin{equation}
\begin{aligned}
    E_{\text{Hartree}}[\tilde{P}] = \frac{1}{2A} \sum_{q_x} & \frac{\tilde{V}^{\text{Hartree}}_{++}(q_x) + \tilde{V}^{\text{Hartree}}_{+-}(q_x)}{2} \Tr\big(\tilde P(q_x)\big) \Tr\big(\tilde P(-q_x)\big) \\
    & + \frac{\tilde{V}^{\text{Hartree}}_{++}(q_x) - \tilde{V}^{\text{Hartree}}_{+-}(q_x)}{2} \Tr\big(\tilde P(q_x) \sigma_z \big) \Tr\big(\tilde P(-q_x) \sigma_z \big) 
\end{aligned}
\end{equation}
The Fock energy has a similar decomposition that reads
\begin{equation}
\begin{aligned}
    E_{\text{Fock}}[\tilde{P}] = -\frac{1}{2A} \sum_{q_x} & \frac{\tilde{V}^{\text{Fock}}_{++}(q_x) + \tilde{V}^{\text{Fock}}_{+-}(q_x)}{2} \Tr\big(\tilde P(q_x)\tilde P(-q_x)\big) \\
    & + \frac{\tilde{V}^{\text{Fock}}_{++}(q_x) - \tilde{V}^{\text{Fock}}_{+-}(q_x)}{2} \Tr\big(\tilde P(q_x) \sigma_z \tilde P(-q_x) \sigma_z \big) 
\end{aligned}
\end{equation}
At the unit total filling $\nu=1$, it is convenient to adopt the pseudospin parameterization
\begin{equation}
    P(k) = \frac{1 + \bm{m}(k) \cdot \bm{\sigma}}{2}\,,\quad
    |\m(k)|=1\,,
\end{equation}
and we have (up to additive constants)
\begin{equation}
\label{eq:Hartree Fock energy pseudospin}
\begin{aligned}
    E_{\text{HF}}[\tilde{\bm{m}}] = \frac{1}{2A} \sum_{q_x} &\frac{\tilde{V}^{\text{Hartree}}_{++}(q_x) - \tilde{V}^{\text{Hartree}}_{+-}(q_x) - \tilde{V}^{\text{Fock}}_{++}(q_x)}{2} \tilde{m}_z(q_x) \tilde{m}_z(-q_x) \\
    & - \frac{\tilde{V}^{\text{Fock}}_{+-}(q_x)}{2} \left(\tilde{m}_x(q_x) \tilde{m}_x(-q_x) + \tilde{m}_y(q_x) \tilde{m}_y(-q_x)\right)\,.
\end{aligned}
\end{equation}
Here $m_z$ measures the charge polarization and is affected by both the Hartree and Fock term, and $m_x$, $m_y$ characterizes the interlayer coherence which comes only from the Fock term.

Assuming that the wave function varies slowly along the $x$ direction we can approximate the Hartree-Fock energy by expanding the interactions at the small momenta $q_x \ell_B \ll 1$. It suffices for our purpose to keep the expansion up to the second order (Note that the linear order of $q_x$ vanishes owing to the reflection symmetry $x \rightarrow -x$):

\begin{equation}
    \tilde{V}^{\text{Hartree}}_{ab}(q_x) = \tilde{V}_{ab}(q_x) e^{-q_x^2\ell_B^2/2} \approx \tilde{V}_{ab}(0) + \left( \frac{\tilde{V}_{ab}''(0)}{2\ell_B^2} - \frac{\tilde{V}_{ab}(0)}{2} \right) q_x^2\ell_B^2
\end{equation}
\begin{equation}
    \begin{aligned}
    \tilde{V}^{\text{Fock}}_{ab}(q_x) &= \frac{1}{N_\phi} \sum_{\q'} \tilde{V}_{ab}(\q') e^{-|\q'|^2\ell_B^2/2} e^{-i q_x q_y'\ell_B^2} \\
    &\approx \frac{1}{N_\phi} \sum_{\q'} \tilde{V}_{ab}(\q') e^{-|\q'|^2\ell_B^2/2} \left( 1 - \frac{q_y'^2 \ell_B^4}{2} q_x^2 \right) \\
    &= \tilde{V}_{ab}^{\text{Fock}}(0) - \left(\frac{1}{N_\phi}\sum_{\q'} \tilde{V}_{ab}(\q') e^{-|\q'|^2\ell_B^2/2} \frac{|\q'|^2 \ell_B^2}{4}\right)q_x^2\ell_B^2
    \end{aligned}
\end{equation}

The leading term penalizes the charge polarization
\begin{equation}
    \frac{\tilde{\beta}}{2} \sum_x m_z^2(x)\,,\quad \tilde{\beta} = \frac{\tilde{V}^{\text{Hartree}}_{++}(0) - \tilde{V}^{\text{Hartree}}_{+-}(0) - \tilde{V}^{\text{Fock}}_{++}(0) + \tilde{V}^{\text{Fock}}_{+-}(0)}{2\times 2\pi \ell_B^2},
\end{equation}
and the next non-vanishing term produces stiffness 
\begin{equation}
    E_{\text{HF}}^{(2)}[\bm{m}] = \frac{\tilde{\rho}_z}{2}  \sum_x \big( \partial_x m_z(x) \big)^2 + \frac{\tilde{\rho}_{xy}}{2}  \sum_x \big( \partial_x m_x(x) \big)^2 + \big( \partial_x m_y(x) \big)^2
\end{equation}
where
\begin{equation}
\begin{aligned}
    \tilde{\rho}_{z} =& \frac{\tilde{V}_{++}''(0) - \tilde{V}_{+-}''(0)}{4\times 2\pi \ell_B^2} - \frac{\tilde{V}_{++}(0) - \tilde{V}_{+-}(0)}{4 \times 2\pi \ell_B^2}\ell_B^2 + \frac{1}{8A} \sum_{\q'} \tilde{V}_{++}(\q') e^{-|\q'|^2\ell_B^2/2} |\q'|^2 \ell_B^2 \\
    \tilde{\rho}_{xy} =& \frac{1}{8A} \sum_{\q'} \tilde{V}_{+-}(\q') e^{-|\q'|^2\ell_B^2/2} |\q'|^2 \ell_B^2
\end{aligned}
\end{equation}
Note that $\tilde{\rho}_z$ here differs from that in Ref.~\cite{moon199513SpontaneousInterlayer} but should be the correct one to use.
Let us also provide details of the Fourier transformation,
$$
\begin{aligned}
\sum_{q_x} q_x^2 \tilde{m}_\alpha(q_x) \tilde{m}_\alpha(-q_x) = & \sum_{q_x} \sum_{k,k'} q_x^2 e^{-iq_x(k-k')\ell_B^2} m_\alpha(k) m_\alpha(k') \\
=& \frac{L_xL_y}{(2\pi)^2} \sum_{k'} \int dq_x dk\, q_x^2 e^{-iq_x(k-k')\ell_B^2} m_\alpha(k) m_\alpha(k') \\
=& \frac{L_xL_y}{(2\pi)^2} \sum_{x'} \int dq_x \frac{dx}{\ell_B^2}\, q_x^2 e^{-iq_x(x - x')} m_\alpha(x) m_\alpha(x') \qquad (k = x/\ell_B^2, k' = x'/\ell_B^2)
\end{aligned}
$$
Note that we define $m_\alpha(x) = m_\alpha(k)$ in the last step so that we still have the normalization $m^2(x) = 1$. Then we have
$$
\begin{aligned}
    \sum_{q_x} q_x^2 \tilde{m}_\alpha(q_x) \tilde{m}_\alpha(-q_x) =&   \frac{L_xL_y}{(2\pi)^2 \ell_B^2} \sum_{x'} \int dq_x dx\, \big(\partial_x^2 e^{-iq_x(x - x')} \big) m_\alpha(x) m_\alpha(x') \\
    =& \frac{L_xL_y}{2\pi \ell_B^2} \sum_{x} \big(\partial_x^2 m_\alpha(x) \big)  m_\alpha(x)
\end{aligned}
$$
When taking the continuum limit we do
$$
    \sum_x \rightarrow \frac{L_y}{2\pi \ell_B^2} \int dx
$$
Thus we have
\begin{equation}
    E_{\text{HF}}[\m] \approx \frac{L_y}{2\pi \ell_B^2} \int \mathrm{d}x\ \frac{\tilde{\rho}_{xy}}{2} \Big( (\partial_x m_x)^2+(\partial_x m_y)^2 \Big)
    + \frac{\tilde{\rho}_z}{2}(\partial_x m_z)^2 + \frac{\tilde{\beta}}{2} m_z^2.
    \label{eq:NLsM}
\end{equation}
We then arrive at the NLsM in main text with $(\beta,\rho_{xy},\rho_{z})=(\tilde{\beta},\tilde{\rho}_{xy},\tilde{\rho}_z)/2\pi\lb^2$. Note that this factor equals the flux density $n_0=N_\phi/A = (2\pi\lb^2)^{-1}$.

Finally, we add the single-particle potential $U_a(\r)$ to our non-linear sigma model, which is more convenient to represent in real space
\begin{equation}
    E_{\text{p}}[\m] = L_yn_0\int dx (U_+(x)m_+ + U_-(x)m_-)=L_y\int dx(\mu(x)-\Delta(x)m_z)
\end{equation}
with
\begin{equation}
    \mu(x)=n_0\frac{U_+(x)+U_-(x)}{2}, \Delta(x) = -n_0 \frac{U_+(x)-U_-(x)}{2}
\end{equation}
We then arrive at the NLsM that we used in the main text:
$E[\m]=E_{\text{HF}}[\m] + E_{\text{p}}[\m] $.

\begin{figure}
\centering
\begin{tikzpicture}[scale = 0.8, baseline={(current bounding box.center)}]
\draw[thick] (0,0) -- (5,0);
\draw[thick] (0,0.6) -- (5,0.6);
\fill[gray!50] (0,1.45) rectangle (5,1.6);
\fill[gray!50] (0,-0.85) rectangle (5,-1);
\draw[>=stealth, <->, thick] (-0.2,-1) --node[left]{$2d_g$} (-0.2, 1.6);
\draw[>=stealth, <->, thick] (5.2,0) -- node[right]{$d_\ell$} (5.2,0.6);
\end{tikzpicture}
\caption{\label{fig:appendixdevice} Geometry of the device that realizes the bilayer quantum Hall system.}
\end{figure}
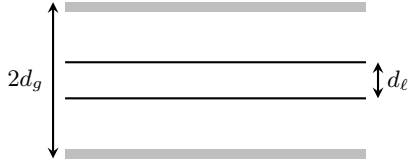

\subsection{Parameters in Non-linear sigma model}
\label{app:NLsM_parameters}
Now we specify interaction $V(q)$ and potential $U(q)$ according to our device and discuss how the NLsM parameters $\beta$ $\rho_{xy}$, $\rho_z$, and $\Delta$ depend on the device geometry.
For convenience we will set $\lb=1$ and $E_c=e^2/(4\pi \epsilon\lb)=1$, and omit the tilde marks tied to $q$-quantities below.

As depicted in \figref{fig:appendixdevice}, we place the active electron in bilayers at $z=\pm d_\ell/2$ and apply electrostatic potentials $\phi_\pm(q)$ to two graphite gates at $z=\pm d_g$. 
The gates have two effects: they imprint external potentials in each layer
\begin{equation}
    U_\pm (q) =\sum_{a= \pm} e\phi_a(q) \frac{\sinh (q(d_g \pm \frac{a d_\ell}{2}))}{\sinh (q d_g)} \frac{1}{2 \cosh (q d_g)},
\end{equation}
and partially screen the Coulomb interaction between electrons
\begin{equation}
    V_{+\pm}(q) = \frac{4\pi}{q} \frac{\sinh(q(d_g \pm \frac{d_\ell}{2}) \sinh(q(d_g-\frac{d_\ell}{2})) }{\sinh (2 q d_g)}.
\end{equation}
As seen in the previous section, the interaction often appears in the following linear combinations
\begin{equation}
\begin{aligned}
    V_0(q) &= \frac{V_{++}(q)+V_{+-}(q)}{2} = \frac{2\pi}{q}\frac{\cosh(qd_\ell/2)\sinh(q(d_g-d_\ell/2))}{\cosh(qd_g)}\\
    V_z(q) &= \frac{V_{++}(q)-V_{+-}(q)}{2} = \frac{2\pi}{q}\frac{\sinh(qd_\ell/2)\sinh(q(d_g-d_\ell/2))}{\sinh(qd_g)}
\end{aligned}
\end{equation}
Below we will derive the NLsM parameters from these analytic expressions and focus on the experimentally relevant regime $d_g\gg 1, d_\ell \ll 1$.

We first analyze the capacitance energy parameter
$$
    \beta = \frac{1}{2\pi}\int_0^{\infty} \frac{dq}{2\pi} \,q e^{-q^2/2} (V_z(0)-V_z(q)) \approx \frac{V_z(0)}{4\pi^2}+\frac{1}{4\pi}
    \int_0^\infty dq e^{-q^2/2}(1-e^{-q d_\ell})\,,
$$
where we have used the approximation $V_z(q)\approx\frac{2\pi}{q}\sinh(qd_\ell/2)e^{-qd_\ell/2}=\frac{\pi}{q}(1-e^{-qd_\ell})$. 
Assuming a small interlayer distance $d_\ell\rightarrow 0$, we can expand the integrand with respect to $d_\ell$ and have
\begin{equation*}
\begin{aligned}
    \beta &\approx \frac{V_z(0)}{4\pi^2} +
    \frac{1}{4\pi}\int_0^\infty dq e^{-q^2/2}\left(-q d_\ell+\frac{q^2d_\ell^2}{2}+...\right)\\
    &= \frac{d_\ell}{4\pi}\left(1-\frac{d_\ell}{2d_g}\right) + \frac{1}{4\pi}\left(-d_\ell+\sqrt{\frac{\pi}{8}}d_\ell^2+...\right)
\end{aligned}
\end{equation*}
We thus find the leading order expansion of $\beta$ in terms of $d_\ell$ and $d_g$:
\begin{equation}
    \beta \approx \frac{d_\ell^2}{8\sqrt{2\pi}}\left(1-\sqrt{\frac{2}{\pi}}d_g^{-1}\right)
    \label{eq:beta}
\end{equation}
The fact that $\beta$ has a quadratic dependence on the interlayer distance $d_\ell$ is important for the single split gate to work in a large parameter regime.

Then we analyze the stiffness parameter $\rho_z$ and $\rho_{xy}$. We first compute
$$
    \rho_z =\frac{1}{8\pi^2}(V_z''(0)-V_z(0))+ \frac{1}{16\pi}\int_0^\infty \frac{dq}{2\pi}e^{-q^2/2}q^3V_{++}(q).
$$
The first term is determined by $V_z(0)$ and $V''_z(0)\approx \frac{\pi}{3}d_\ell^3$. (Note that $d_g\rightarrow\infty$ is taken before computing $q\rightarrow0$) The second term can be approximated given  $V_{++}(q)\approx\frac{2\pi}{q}(1-2e^{-2qd_g}+2e^{-4qd_g}+...)$:
\begin{equation*}
    \int_0^\infty dq e^{-q^2/2}q^2(1-2e^{-2qd_g}+2e^{-4qd_g}+...) \xrightarrow{d_g\rightarrow \infty} \sqrt{\frac{\pi}{2}}-2\int_0^\infty dq q^2(e^{-2qd_g}-e^{-4qd_g}+...)=\sqrt{\frac{\pi}{2}}+O(d_g^{-3})
\end{equation*}
Therefore to the leading order we have
\begin{equation}
    \rho_z \approx \frac{1}{16\sqrt{2\pi}}-\frac{1}{8\pi}d_\ell\left(1-\frac{d_\ell}{2d_g}\right)+\frac{1}{24\pi}d_\ell^3,
    \label{eq:rho_z}
\end{equation}
which is valid as long as we take $d_\ell\rightarrow 0$ before $d_g\rightarrow \infty$. 
We can similarly compute
$$
    \rho_{xy} = \frac{1}{16\pi}\int_0^\infty \frac{dq}{2\pi}e^{-q^2/2}q^3V_{+-}(q).
$$
Given $V_{+-}(q)\approx\frac{2\pi}{q}(e^{-qd_\ell}-2e^{-2qd_g}+2e^{-4qd_g}\cosh(qd_\ell)+...)$, we have
\begin{equation}
\begin{aligned}
    \rho_{xy} &\xrightarrow{d_g\rightarrow \infty} \frac{1}{16\pi}\int_0^\infty dq e^{-q^2/2}q^2e^{-qd_\ell} + O(d_g^{-3}) \xrightarrow{d_\ell\rightarrow 0} \frac{1}{16\pi}\int_0^\infty dq e^{-q^2/2}q^2(1-qd_\ell+...) + O(d_g^{-3})\\
    &\approx \frac{1}{16\sqrt{2\pi}} -\frac{1}{8\pi}d_\ell
    \label{eq:rho_xy}
\end{aligned}
\end{equation}
Thus, we conclude that $\rho_{xy}=\rho_{z}=\frac{E_c}{16\sqrt{2\pi}}$ at $d_\ell=0$ and is reduced at the order of $O(d_\ell)$ for $d_\ell\neq 0$.

Finally, we analyze the displacement potential
\begin{equation}
    \Delta(q) = -n_0 \frac{U_+(q)-U_-(q)}{2}=-e(\phi_+(q)-\phi_-(q))\frac{\sinh(qd_\ell/2)}{4\pi\sinh(qd_g)}.
\end{equation}
For our single split gate setup: $\phi_+(|x|<L_J/2) =\mu/e+\phi_t$ and $\phi_+(|x|>L_J/2)=\phi_-(x)=\mu/e$, we have $\phi_+(x)-\phi_-(x)=\phi_t(\Theta(x+L_J/2)-\Theta(x-L_J/2))$, leading to
\begin{equation}
    \phi_+(q)-\phi_-(q)=\int_{-L_J/2}^{L_J/2}dx e^{iqx}\phi_t =\phi_t \frac{\sin(qL_J)}{qL_J}
\end{equation}
Therefore
\begin{equation}
    \Delta(q) = -e\phi_t\frac{\sin(qL_J)}{qL_J}\frac{\sinh(qd_\ell/2)}{4\pi\sinh(qd_g)}
\end{equation}
We can then evaluate $\Delta(x)$ using the convolution formula, with $\mathcal{F}^{-1}$ denoting the inverse Fourier transform:
\begin{equation}
    \Delta(x) = -\frac{e\phi_t}{4\pi}\int_{-L_J/2}^{L_J/2}dx'\mathcal{F}^{-1}\left[\frac{\sinh(qd_\ell/2)}{\sinh(qd_g)}\right](x-x')
\end{equation}
If we approximate $d_\ell\ll 1$, we have
\begin{equation}
    \mathcal{F}^{-1}\left[\frac{\sinh(qd_\ell/2)}{\sinh(qd_g)}\right]\approx \mathcal{F}^{-1}\left[\frac{qd_\ell/2}{\sinh(qd_g)}\right]=\frac{\pi d_\ell}{8d_g^2\cosh^{2}\left(\frac{\pi x}{2d_g}\right)}
\end{equation}
then we have
\begin{equation}
    \Delta(x) = -\frac{e\phi_t d_\ell}{4\pi}\int_{-L_J/2+x}^{L_J/2+x}dx'\frac{\pi}{8d_g^2\cosh^{2}\left(\frac{\pi x'}{2d_g}\right)}=-\frac{e\phi_t d_\ell}{16\pi d_g}\left.\tanh\left(\frac{\pi x}{2d_g}\right)\right|_{-L_J/2+x}^{L_J/2+x}
\end{equation}
For $|x\pm L_J/2|\ll d_g$, we obtain the result consistent with the uniform potential $\Delta(x)\approx\Delta_0 =\frac{d_\ell}{4d_g}\frac{e\phi_t}{2\pi}$.

\subsection{Charge gap of the exciton condensate and layer-polarized state and the resulting restrictions on $\phi_t$}
\label{app:charge gap}

In this section, we use the Hartree-Fock Hamiltonian to estimate the charge gap of the exciton condensate ($m_z = 0$) and the layer-polarized state ($m_z \approx 1$), from which we determine the range of allowed $\phi_t$ in the single-split gate setup.

Assuming a uniform ground state, the Hatree-Fock Hamiltonian is
\begin{equation}
\begin{gathered}
    H_\text{HF} = \sum_{ab,k} (V^H_{ab} P_{bb} + U_a) c_{k,a}^\dag c_{k,a} - \sum_{ab,k} V^F_{ab}P_{ba}c_{k,a}^\dag c_{k,b}, \\
    V^H_{ab} \equiv \frac{\tilde{V}_{ab}(0)}{2\pi\ell_B^2},\ \ 
    V^F_{ab} \equiv \int\frac{d^2q}{(2\pi)^2} \tilde{V}_{ab}(\bm{q})e^{-|\bm{q}|^2\ell_B^2/2}
\end{gathered}
\end{equation}
where $a=(l_a,\tau_a), b=(l_b,\tau_b)$, $l_a = \pm$ denotes the up/down layer, $\tau$ denotes valley and spin flavor, $U_a$ denotes the bare energy of each flavor, the screened Coulomb interaction $\tilde{V}_{ab}$ is independent of $\tau_a$ and $\tau_b$.
Note that it takes the same form for different Landau level orbitals $k$ takes.
Therefore, we only keep track of the flavor indices of fermion operators to simplify our notation.

For the exciton condensate phase $P_{ab} = \delta_{\tau_a, \tau_b}\delta_{\tau_a, K\uparrow}/2$, the Hartree-Fock Hamiltonian per orbital is
\begin{equation}
    \sum_{l_a,l_b}\frac{V_{++}^H + V_{+-}^H-V_{++}^F}{2}c^{\dag}_{l_aK\uparrow}c_{k, l_aK\uparrow} -\frac{V_{+-}^F}{2}c^{\dag}_{l_aK\uparrow}\sigma^x_{l_a,l_b}c_{l_bK\uparrow}
    +\sum_{\tau_a\neq K\uparrow}\sum_{l_a}\frac{V_{++}^H + V_{+-}^H}{2}c^{\dag}_{a}c_{a}
\end{equation}
For this solution to be self-consistent, the gate voltage $\phi_0$ must satisfy

\begin{equation}
\begin{aligned}
    \frac{V_{++}^H + V_{+-}^H - V_{++}^F - V_{+-}^F}{2} - e\phi_0 &< 0\\
    \frac{V_{++}^H + V_{+-}^H - V_{++}^F + V_{+-}^F}{2} - e\phi_0 &> 0
\end{aligned}
\label{eq:xoccupied}
\end{equation}
On the other hand, for the layer-polarized state $P_{ab} = \delta_{a, b}\delta_{a,+K\uparrow}$, we have
\begin{align}
    (V_{++}^H-V_{++}^F-U_\Delta)c^{\dag}_{ +K\uparrow}c_{+K\uparrow} + \sum_{\tau_a\neq K\uparrow} (V_{++}^H-U_\Delta)c^{\dag}_{+\tau_a}c_{+\tau_a}
    &+ \sum_{\tau_a}(V_{+-}^H+U_\Delta)c^{\dag}_{-\tau_a}c_{-\tau_a}
\end{align}
where $U_\Delta = \frac{d_\ell}{4d_g}e\phi_t$ is the layer polarization potential. 
For this solution to be self-consistent inside the junction, the average gate voltage of the top middle gate and the bottom gate must satisfy
\begin{align}
    V_{++}^H - V_{++}^F - e(\phi_0+\frac{\phi_t}{2}) - U_\Delta &< 0\\
    V_{++}^H - e(\phi_0+\frac{\phi_t}{2}) - U_\Delta &> 0\label{eq:+notoccupied}\\
    V_{+-}^H - e(\phi_0+\frac{\phi_t}{2}) + U_\Delta &> 0
    \label{eq:-notoccupied},
\end{align}
Note that $V_{++}^H - V_{+-}^H\propto d_\ell$. Consider the case $d_\ell\ll \ell_B$ and $\Delta\propto\beta\sim d_\ell^2$, Eq.~\ref{eq:-notoccupied} implies Eq.~\ref{eq:+notoccupied}.
Thus, $\phi_t$ must satisfy
\begin{equation}
   \frac{2(V_{++}^H - V_{++}^F-e\phi_0)}{1+d_l/2d_g} <\phi_t< \frac{2(V_{+-}^H -e\phi_0)}{1-d_l/2d_g}
\end{equation}
Taking $e\phi_0 = \frac{V_{++}^H + V_{+-}^H - V_{++}^F - V_{+-}^F}{2}$, the lower bound given by Eq.~\ref{eq:xoccupied}, we have
\begin{equation}
    \frac{4\pi\beta}{1+d_l/2d_g}<\phi_t < \frac{2(V_{+-}^F-2\pi\beta)}{1-d_l/2d_g},
    \label{eq:phi_min_max_full}
\end{equation}
where $\beta$ is given by \eqnref{eq:beta}. The upper bound for $\phi_0$ can only be smaller for other allowed values of $\phi_0$. In the limit $d_l \ll \ell_B,d_g$, the upper bound is approximately $2V_{+-}^F$. 

\begin{figure}
    \centering    \includegraphics[width=0.35\textwidth]{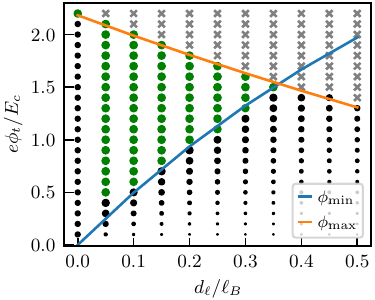}
    \caption{Hartree-Fock result of $m_z$ and filling $\nu$ at different $d_\ell$ and $\phi_t$, with $d_g=5\lb$, $L_x = 9L_y = 48\sqrt{2\pi}\lb$ and $L_J=L_x/2$. Crosses represent gapless states ($\nu>1$), and circles represent gapped states ($\nu=1$). The size of the circle is proportional to $m_z^2$, with $m_z^2>0.99$ cases colored by green. The two curves correspond to $\phi_\mathrm{min}<\phi_t<\phi_\mathrm{max}$ in \eqnref{eq:phi_min_max_approx}, which shows a decent agreement between the analytic results and the HF numerics.}
    \label{fig:Hartree-Fock}
\end{figure}

In addition, according to the non-linear sigma model, the layer-polarized state is the true ground state only when $\Delta = \frac{d_\ell}{4d_g}\frac{e\phi_t}{2\pi} > \beta$. Thus, the allowed range of $\phi_t$ is
\begin{equation}
    8\pi d_g\beta/d_l< e\phi_t < 2 V_{+-}^F
    \label{eq:phi_min_max_approx}
\end{equation}
For the dual gate configuration, the parameters on the two sides of the inequalities are
\begin{equation}
    \beta \approx \frac{d_\ell^2}{8\sqrt{2\pi}}\left(1-\sqrt{\frac{2}{\pi}}d_g^{-1}\right)\,,\quad
    V_{+-}^F = \sqrt{\pi/2}-d_\ell-(\ln2)d_g^{-1},
\end{equation}
therefore $\phi_\mathrm{min}=8\pi d_g\beta/d_l\approx \sqrt{\pi/2}\,d_g d_\ell$, and $\phi_\mathrm{max}= 2V_{+-}^F\approx \sqrt{2\pi}$.
As a result, $\phi_t$ has a smaller range of tunability at larger $d_\ell$ (smaller $\lb$), and above a critical value $d_\ell d_g\approx 2\lb^2$ no $\phi_t$ can satisfy the constraints.
The result can be verified with Hartree-Fock numerics, as shown in Fig.\,\ref{fig:Hartree-Fock}.

Aside from point-like excitations, the exciton condensate state also admits charged excitations with extended pseudospin textures known as bimerons \cite{moon199513SpontaneousInterlayer}. For the $m_z=0$ state, each bimeron comprises of two merons with opposite vortices in the $m_{xy}$ texture, with their core polarized in either one of the two layers. Each meron carries a charge $\pm e/2$, so the bimeron carries a charge $\pm 1$ or $0$. For $d_\ell/\lb\sim 0.2$, the charged and neutral bimeron energy is estimated to be $0.5V_{+-}^F$ and $0.25V_{+-}^F$ respectively \cite{brey1996ChargedPseudospinTextures}, smaller than the charge gap from uniform Hartree-Fock analysis. Therefore we expect the realistic constraint on $\phi_t$ to be stronger but qualitatively close to our estimate.

\section{Validity of lumped-element description}
\label{app:lumped_element_validity}

The actual device of the Josephson junction comprises three parts: the junction in the middle and two EC bulks on the two sides.
The lumped-element description approximates each part as a locally defined component.
For example, the expression of the Josephson energy $E_J\cos\varphi$ captures the phase difference $\varphi$ across the junction by implicitly ignoring phase variations within each EC bulk.
The neglected contribution can be significant for certain device configurations, especially in the case of strongly-linked junctions. This section revisits the microscopic description of the Josephson junction and discusses when the lumped-element treatment is valid.

Our discussion is based on the NLsM of the Josephson junction \eqnref{eq:LandauGinzburg}, quoted below for the reader's convenience
\begin{equation}
    \frac{E[\bm{m}]}{L_y} = \int \mathrm{d}x\ \frac{\rho_{xy}}{2}\left[(\partial_x m_x)^2+(\partial_x m_y)^2\right] 
    + \frac{\rho_z}{2}(\partial_x m_z)^2 + \frac{\beta}{2} m_z^2-\Delta(x) m_z.
\end{equation}
We place the junction at $0 < x < L_J$ and two EC bulks at $-L_x<x<0$ and $L_J<x<L_J + L_x'$ respectively. 
In principle, both the left and right EC islands contribute to the capacitance and inductance in the lump-element description.
Below we focus on the left EC bulk and fix $\varphi(L_J)=0$ to get rid of the right EC island to simplify our discussion. 
Including the right EC bulk only complicates the calculation but will not quantitatively change the conclusion.

Based on the discussion in the main text, we can approximate the junction energy by $-E_J \cos \varphi(0)$, and assume $m_z\approx 0$ and slow-varying in the left EC islands.
We have
\begin{equation}
\begin{aligned}
    \frac{E[\bm{m}]}{L_y} \approx \int_{-L_x}^0  dx\ \Big( \frac{\rho}{2}(\partial_x\varphi)^2 +\frac{\beta}{2} m_z^2 - \frac{eV(x)}{4\pi} m_z \Big) - \varepsilon_J\cos\varphi(0).
    \label{eq:lumped_element_NLSM}
\end{aligned}
\end{equation}
where $\varepsilon_J=E_J/L_y$ is the lumped-element Josephson energy per unit width. 
The energy reduces to that for a single capacitance $C_J^{-1}\propto\beta$ if we take $\partial_x\varphi=0$.
Below we estimate corrections to the lumped-element approximation due to spatial fluctuations in $\varphi$. We set $\ell_B = 1$ for simplicity.

Specifically, we examine the system's response to an applied driving voltage and analyze how the spatial variation of $\varphi$ modifies the resonant frequency.
To capture such dynamics, a fully quantum mechanical treatment of the problem is required.
We choose the pair of canonical variables to be the phase $\varphi$ and the electron density $\Pi_\varphi = \hbar n = \hbar m_z / 4\pi$, which is consistent with the choice of the main text (for example \eqnref{eq:lumped_element_EOM}). 
The Hamiltonian reads
\begin{equation}
    H = \int_{-L_x}^0 dx\ \Big( \frac{\rho}{2}(\partial_x\varphi)^2 +\frac{8\pi^2\beta}{\hbar^2} \Pi_\varphi^2 - \frac{eV(x,t)}{\hbar} \Pi_\varphi \Big) - \varepsilon_J\cos\varphi(0).
\end{equation}
In the bulk, the equations of motion of $(\varphi,\Pi_\varphi)$ are
\begin{equation}
\left\{
\begin{aligned}
    \partial_t{\varphi}(x)&=\frac{\partial H}{\partial \Pi_\varphi(x)}=\frac{16\pi^2\beta}{\hbar^2}\Pi_\varphi(x)-\frac{e}{\hbar}V(x) \\
    \partial_t{\Pi}_\varphi(x) &= -\frac{\partial H}{\partial \varphi(x)} = \rho \partial_x^2\varphi(x).
\end{aligned}
\right.
\end{equation}
which leads to 
\begin{equation}
    (\partial_t^2 - v^2 \partial_x^2) \varphi = \frac{e}{\hbar}\frac{\partial V(x,t)}{\partial t}\,,\quad v=\frac{4\pi}{\hbar} \sqrt{\rho\beta}\,.
\end{equation}
On the two boundaries $x=0$ and $x=-L_x$, the current conservation law provides the proper boundary conditions
\begin{equation}
    \varepsilon_J \sin\varphi(0) + \rho\partial_x\varphi(0) = 0\,,\quad \rho\partial_x\varphi(-L_x)=0\,.
\end{equation}
Now, we turn on an AC driving voltage $V(x,t)=V_0\cos\omega t$ and consider the following trial solution
\begin{equation}
    \varphi(x,t)=\Re\big[ \big( \varphi_0+A_k\cos (k(x+L_x)) \big)e^{i\omega t} \big]\,,\quad \omega^2 = v^2 k^2\,.
\end{equation}
The boundary condition at $x = -L_J$ is automatically satisfied.
The bulk equation of motion determines $\varphi_0$:
\begin{equation}
    -\omega^2\Re[\varphi_0 e^{i\omega t}] = \frac{e}{\hbar}\Re{V_0 i\omega e^{i\omega t}}\Longrightarrow \varphi_0 = \frac{eV_0}{i\hbar\omega}.
\end{equation}
At the boundary $x = 0$, we assume $|\varphi(0,t)| \ll 1 $ and have a linearized equation
\begin{equation}
    \varepsilon_J\varphi_0 + A_k(\varepsilon_J\cos kL_x-\rho k \sin kL_x) = 0,
\end{equation}
The resonance, when $A_k\rightarrow\infty$, occurs at wave vectors that satisfy the following condition
\begin{equation}
     kL_x\tan kL_x = \varepsilon_J L_x/\rho\,,
\end{equation}
The resonance condition has infinitely many solutions, but only the lowest resonant solution $k_J$ is relevant at low probe frequency.

If we assume $k_JL_x\ll 1$, we can approximate $\tan(k_JL_x)\approx k_JL_x$. Then we have $k_J = \sqrt{\varepsilon_J/\rho L_x}$ and
\begin{equation}
    \omega_J^2 = v^2 k_J^2 = \big( \frac{4\pi}{\hbar} \big)^2 \frac{\beta}{L_x} \frac{1}{\varepsilon_J^{-1}}\,.
    \label{eq:app_Josephson_frequency}
\end{equation}
Recalling that $E_J = \varepsilon_J L_y$ and $C_J^{-1}=16\pi^2\beta/L_x L_y$, this recovers the lumped-element result $\omega_J = e\sqrt{E_J/C_J}/\hbar$ in \eqnref{eq:Josephson_frequency}. Meanwhile, a more accurate solution of $k_J$ is given by the series expansion of $\varepsilon_J L_x/\rho$:
\begin{equation}
    k_J = \sqrt{\frac{\varepsilon_J}{\rho L_x}} \Big( 1-\frac{\varepsilon_J L_x}{6\rho}+\frac{11}{360}\Big(\frac{\varepsilon_J L_x}{\rho}\Big)^2+\ldots\Big).
    \label{eq:kj series}
\end{equation}
Physically, $\varepsilon_J L_x/\rho$ can be interpreted as the ratio of the edge inductance from Josephson energy $
L_{\text{edge}} \sim \varepsilon_J^{-1}$ and bulk kinetic inductance $L_{\text{bulk}} \sim L_x/\rho$.
To see its contribution more clearly, we express the corresponding resonance frequency as in \eqnref{eq:app_Josephson_frequency}
\begin{equation}
    \omega_J^2 = \big( \frac{4\pi}{\hbar} \big)^2 \frac{\beta}{L_x} \frac{1}{\varepsilon_J^{-1} + \frac{L_x}{3\rho} + \ldots}\,,
\end{equation}
in which the edge inductance $(\varepsilon_J^{-1})$ is substituted with the sum of edge and bulk inductance in series $(\varepsilon_J^{-1} + \frac{L_x}{3\rho})$.
We also note that including contribution from the right EC island will merely add quantitative modifications to the capacitance $\beta/L_x$ on the numerator and the inductance on the denominator. 
Therefore, the accurate $\omega_J$ should account for the inductance of both the edge and the bulk, and the lumped-element description is valid only when the bulk inductance is negligible, or quantitatively, 
\begin{equation}
    k_J L_x=\omega_J L_x/v \ll 1 \ \Leftrightarrow \ \varepsilon_J L_x/\rho\ll 1
\end{equation}
The two equivalent conditions cover different perspectives: the second from the strength of Josephson coupling, and the first from the resulting resonant frequency. The same conclusion can be applied to the right EC bulk, leading to similar conditions: $\omega_J L'_x/v \ll 1\,\Leftrightarrow\, \varepsilon_J L'_x/\rho\ll 1$.

Finally, we estimate when the lumped-element treatment is justified in realistic devices.
From previous sections we have $v \sim \sqrt{\rho\beta}/\hbar \sim E_c d_\ell/\hbar \sim \alpha_\text{BN}  d_\ell c$ (with $\alpha_\text{BN} = \frac{1}{4.4}\frac{1}{137}$ the fine-structure in hBN and $c$ the velocity of light) and $\varepsilon_J\sim \sqrt{\rho\beta}\exp(-L_J\sqrt{\beta/\rho})\sim E_c d_\ell \exp(-\sqrt{2} L_J d_\ell)$ (for weakly-linked junction). If we take $L_x= \SI{5}{\micro\meter}$ and $d_\ell = 0.2\ell_B=\SI{2}{\nano\meter}$ $(B=\SI{6.5}{T})$, the conditions leads to
\begin{equation}
    \nu_J\ll\SI{20}{GHz} \ \Leftrightarrow\  L_J>\SI{240}{nm}\ (\mathrm{with}\ \varepsilon_J L_x/\rho=0.1)
\end{equation}
Note that resonance can still be observed in experiments even when deviating from the above conditions, but needs to be corrected accordingly to match the theoretical description.

\end{document}